\documentclass{phbauth}
\usepackage{graphicx}

\begin{document}

\begin{frontmatter}

\title{Anomalous scaling dimensions and 
critical points in type-II superconductors}

\author[address1]{Asle Sudb{\o}\thanksref{thank1}} 
\author[address1]{Anh Kiet Nguyen} and
\author[address1]{Joakim Hove}

\address[address1]{Department of Physics, Norwegian University of Science
and Technology, N-7491 Trondheim, Norway}
\thanks[thank1]{Corresponding author. FAX: +47 73 59 77 10. 
                                      E-mail: asudbo@phys.ntnu.no}

\begin{abstract}
  The existence of a {\it stable critical point}, separate from the
  Gaussian and XY critical points, of the Ginzburg-Landau theory for
  superconductors, is demonstrated by direct extraction via
  Monte-Carlo simulations, of a negative anomalous dimension $\eta_{\phi}$
  of a complex
  scalar field $\phi$  forming a dual description of a neutral superfluid.
  The dual of the neutral superfluid is isomorphic to a charged
  superfluid coupled to a massless gauge-field. The anomalous scaling
  dimension of the superfluid order-field is positive, while we find
  that the anomalous dimension of the dual field is negative. The dual
  gauge-field does not decouple from the dual complex matter-field at
  the critical point.  {\it These two critical theories represent
    separate fixed points.}  The physical meaning of a negative $\eta_{\phi}$
  is that the vortex-loop tangle of the superfluid at the critical
  point fills space {\it more} efficiently than random walkers,  {\it
    without collapsing}. This is due to the presence of the massless
  dual gauge-field, and the resulting long-ranged {\it vectorial}
  Biot-Savart interaction between vortex-loop segments, which is a
  relevant perturbation to the steric $|\psi|^4$ repulsion term.
  Hence, the critical dual theory is not in the universality class of
  the $|\psi|^4$-theory.
\end{abstract}

\end{frontmatter}

A charged superfluid (spin-singlet superconductor) is described by the
Ginzburg-Landau theory of a complex matter field $\psi$ coupled to a
massless gauge-field ${\bf A}$, with coupling constant $2e$ and a
local gauge-symmetry; here $e$ is the electron charge. In $3D$, the dual 
of this theory is given by a dual complex matter-field $\phi$ coupled to a 
massive dual gauge-field
${\bf h}$ with mass $e$, exhibiting a global $U(1)$-symmetry due to
the mass of ${\bf h}$ \cite{Kleinert:89}. It renders the critical
point of the theory in the universality class of the
$|\phi|^4$-theory. In renormalization group sense, the {\it
  short-range} interaction mediated by the massive ${\bf h}$, is an
irrelevant perturbation to the steric repulsion stemming from the
$|\psi|^4$-term \cite{Dasgupta:81}.  However, in the limit $e \to 0$,
the dual gauge-field becomes massless, and the symmetry of the problem
changes to a local gauge-symmetry. This profoundly changes the
physics.

In this case, the $|\psi|^4$-theory of the $3D$ superfluid,
$H_{\psi}=|\partial_i \psi|^2+m_{\psi}^2 |\psi|^2+(u_{\psi}/2)
|\psi|^4$ has a dual theory given by $H_{\phi,{\bf h}}=|(\partial_i-i
g h_i) \phi|^2 + m_{\phi}^2 |\phi|^2 + (u_{\phi}/2) |\phi|^4 + (1/2)
({\bf\nabla} {\bf \times} {\bf h} )^2$ \cite{Kleinert:89}.  The
coupling constant $g$ is given by $g=2 \pi m_h/e$, where $m_h = e
\omega$, where $\omega$ is the {\it amplitude} of the order-field
$\psi$. Thus, the dual of the neutral superfluid is isomorphic to the
Ginzburg-Landau theory of a charged superfluid coupled to a massless
(dual) gauge-field ${\bf h}$. In Ref. \cite{Kiometzis:93} it is argued that as $ T \to
T_c^{-}$, $\omega \to 0$. This would imply that at and above $T_c$,
${\bf h}$ decouples from  $\phi$. Thus, the critical
{\it dual} theory would be that of a $|\phi|^4$-theory, i.e. the
superfluid and its dual theory would be {\it self-dual}.

This is a mean-field argument. At the true $T_c$,
$\omega > 0$, and hence the dual gauge-field
does not decouple from $\phi$ as $T \to T_c^{-}$.  At the true
critical point, the dual of the $|\psi|^4$-theory is not a
$|\phi|^4$-theory, in particular the symmetries of $H_{\psi}$ and 
$H_{\phi,{\bf h}}$, are
different. We thus expect the critical exponents of these two theories
to be different, describing two separate critical points. This has
been checked by large-scale Monte-Carlo simulations \cite{Nguyen:99}.

The main point of this short communication is the following. Direct
evaluation of the anomalous dimension of the matter field $\psi$,
yields $\eta_{\psi} = 0.038$ \cite{Hasenbusch:99}. In
contrast, the anomalous dimension of the dual field $\phi$, is given
by $\eta_{\phi}=-0.18 \pm 0.07$ \cite{Nguyen:99}. The latter result is
obtained in our Monte-Carlo simulations by analysing the statistics of
vortex-loops in the $3DXY$-model, using a mapping to the dual theory.
Therefore, we {\it demonstrate} that the critical behavior of the
superfluid, which is a $|\psi|^4$-theory, is in a different
universality class than the critical behavior of the dual theory. 
The results of Ref. \cite{Herbut:96} thus rest on a firm theoretical
footing, in that the existence of a novel charged fixed point need not
be {\it assumed}, it is {\it demonstrated}.

The results for the vortex-loop distribution function $D(p)$ as a
function of vortex-loop perimeter $p$ are shown in Fig. 1
\cite{Nguyen:99}. We fit $D(p)$ to the form $D(p) = A p^{-\alpha} ~
\exp(-\varepsilon(T)/k_B T)$ where $T$ is temperature, $\alpha \approx
2.5$, and $\varepsilon(T)$ is the line tension of the thermally
generated vortex-loops, the topological defects destroying 
superfluid order \cite{Nguyen:99}.

We fit $T$-dependence of $\varepsilon(T)$ to the
form $\varepsilon(T) \sim |T-T_c|^{\gamma}$ \cite{Nguyen:99}; 
$\gamma$ is identified as the susceptibility exponent 
$\gamma_{\phi}$ of the
dual field $\phi$ \cite{Nguyen:99}. By the scaling law $\gamma_{\phi}
= \nu_{\phi} (2-\eta_{\phi})$ and the observation that $\nu_{\phi} =
\nu_{\psi}$ \cite{Herbut:96,Nguyen:99}, where $\nu_{\psi}=0.672$
\cite{Hasenbusch:99}, {\it we extract $\eta_{\phi}$ from the
  line-tension of the vortex-loops of the $3DXY$-model}. We find
$\gamma_{\phi} = 1.45 \pm 0.05$, and hence $\eta_{\phi} = -0.18 \pm
0.07$.  This  should
be compared to $\eta_{\psi} = 0.038$ \cite{Hasenbusch:99}. Thus, the
interaction mediated by the massless gauge-field ${\bf h}$ is a relevant
perturbation to the non-trivial fixed-point of the $|\phi|^4$-theory, 
in renormalization group sense, 

\begin{figure}[h]
  \begin{picture}(0,165)(0,0)
     \put(-30,-188)
         {\includegraphics[angle=0,scale=0.49]
         {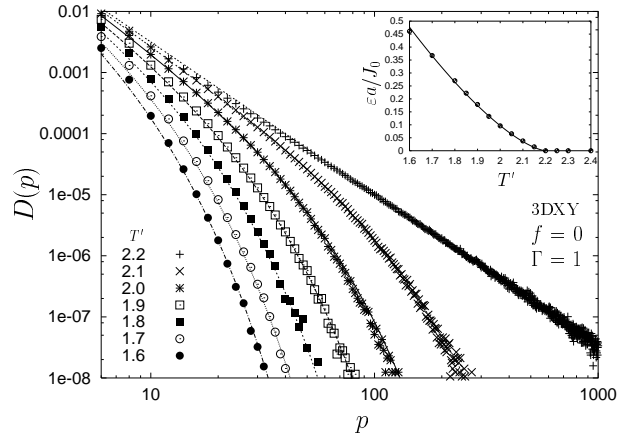}}
 \end{picture}
\caption{Vortex-loop distribution function $D(p)$ as a function of 
  perimeter $p$, at various temperatures. Inset shows vortex-line
  tension $\varepsilon(T)$ close to $T_c$.}
\label{figurename}
\end{figure}

We have established a difference {\it in sign} of $\eta_{\psi}$ and
$\eta_{\phi}$ via Monte-Carlo simulations of the vortex-loop gas of
the neutral superfluid at the critical point. {\it Hence, {\bf h}
does not decouple from $\phi$ at $T_c$.}
The dual
theory is isomorphic to a superconductor coupled to an electromagnetic
gauge-field. Thus, we {\it establish} the existence of a new stable 
charged fixed point of the Ginzburg-Landau theory, of the type 
proposed in Ref.\cite{Herbut:96}.

This dual description is also useful in investigating broken
symmetries in the {\it vortex-liquid phase} of type-II
superconductors in magnetic fields \cite{Tesanovic:99}.

\end{document}